# The Chemical Evolution of Narrow Emission Line Galaxies: the Key to their Formation Processes

J. P. Torres-Papaqui*, R. Coziol*, R. A. Ortega-Minakata*

**ABSTRACT**

Using the largest sample of narrow emission line galaxies available so far, we show that their spectral characteristics are correlated with different physical parameters, like the chemical abundances, the morphologies, the masses of the bulge and the mean stellar age of the stellar populations of the host galaxies. It suggests that the spectral variations observed in standard spectroscopic diagnostic diagrams are not due solely to variations of ionization parameters or structures but reflect also the chemical evolution of the galaxies, which in turn can be explained by different galaxy formation processes.

**RESUMEN**

Utilizando la mayor muestra de galaxias con líneas de emisión angostas disponible hasta el momento, se muestra que sus características espectrales están correlacionadas con diferentes parámetros físicos, como las abundancias químicas, las morfologías, las masas del bulbo, y la edad estelar promedio de las poblaciones estelares de la galaxia anfitriona. Por lo tanto, se sugiere que las variaciones espectrales observadas en diagramas de diagnóstico estándares no se deben únicamente a las variaciones de los parámetros o las estructuras de ionización, sino que reflejan también la evolución química de las galaxias, relacionada con diferentes procesos de formación.

## INTRODUCTION

With the last Data Release (DR 7, Abazajian *et al.*, 2009) of the Sloan Digital Sky Survey (SDSS) (York *et al.*, 2000; Stoughton *et al.*, 2002), the astronomical community has for the first time the possibility of studying in great details a huge sample of nearby galaxies-defined here as galaxies with redshift $z < 0.25$-increasing consequently our hope to understand better these systems. One surprising result of the SDDS project is that almost all (~ 97 %) of the galaxies included in the spectral survey turn out to show some kind of activity under the form of narrow emission lines. This result does not seem to be due to an observational bias (although the fibers separation of the spectroscope does introduce some incompleteness in very dense environments of galaxies), but most possibly reflects a real physical condition of the nearby universe. In terms of structures, for example, it is known that passive galaxies (galaxies without emission lines) are more numerous in dense systems like clusters of galaxies, which are much less common than the standard environment of galaxies at low redshift, that we call loosely the field, and which is formed of long filaments of groups of galaxies of various relatively low densities in galaxies (possibly similar to our own local group, or even less dense like loose groups).

This high amount of Narrow Emission Line Galaxies (NELGs) is in some way fortunate because of the numerous tools that were developed over the years to study such galaxies. More specifically, it was shown that it is possible to distinguish between two main sources of ionization of the gas, separating

* Departamento de Astronomía, División de Ciencias Naturales y Exactas, Universidad de Guanajuato. Apartado Postal 144, C. P.: 36000. Guanajuato, Gto., México. Correos electrónicos: papaqui@astro.ugto.mx, rcoziol@astro.ugto.mx, rene@astro.ugto.mx

"normal" star forming activity from accretion of matter onto a super massive black hole (SMBH) in the center of galaxies, by comparing different emission line ratios in special diagnostic diagrams (Baldwin *et al.*, 1981, Veilleux and Osterbrock 1987). These seminal studies were followed by intense discussion on what is the best way to classify NELGs in the most objective manner, which resulted in the adoption of semi-empirical classification criteria (Kewley *et al.*, 2001; Kauffmann *et al.*, 2003). Applying these criteria it is found that the majority of the SDSS NELGs fall into the Star Formation Galaxies (SFGs) class. However, the distinction between SFGs and the Active Galactic Nuclei (AGNs), where the gas is assumed to be ionized by accretion of matter onto a SMBH, is not as clear as previously thought, and a buffer zone was introduced between these two classes. The concept of Transition type Objects (TOs)-that is, galaxies that show mixed spectral characteristics, suggesting they are both SFGs and AGNs-has won a lot of credence over the years and this class is now included as a standard in NELGs classification studies. On the other hand, the adoption of a new class has introduced the difficulty of understanding the different nature and possible relations between all these types of galaxies.

Even the NELGs falling on the AGN class presents some ambiguities. Historically, only the narrow line galaxies known as Seyfert 2 (S2) were recognized as AGNs (*e.g.* Osterbrock 1989). However, and as many studies have shown since then, the S2 does not form the bulk of narrow line AGNs (NLAGNs), but LINERs do (where LINER stands for Low Ionization Nuclear Emission-line Region; Heckman 1980; Coziol 1996, Kewley *et al.*, 2006). The question is and remains: on what physical basis can we distinguish between these two kinds of AGNs? In general, is it possible to distinguish in the spectra of the NELGs intrinsic characteristics that guarantee the different nature of activity of the galaxies?

In the literature, the problem is usually approached under the form of discussion about different ionization models. In particular, it is assumed that models for SFGs do not apply to AGNs, implying that different photo-ionization codes are necessary for reproducing the line ratios of these more "exotic" objects (one important example of such code is CLOUDY; Ferland *et al.*, 1998). This is especially true for chemical studies. The present state of facts suggest that although we can determine relatively easily the abundances of SFGs (*e.g.* Thurston *et al.*, 1996; van Zee *et al.*, 1998; Coziol *et al.*, 1999; Thuan *et al.*, 2010), mainly the ratios of oxygen to hydrogen, O/H (or [O/H] when given relative to the solar ratio), of nitrogen to oxygen, N/O, or nitrogen to hydrogen, N/H (in astronomy all elements heavier than helium are considerate as metal, and the ratio [O/H] is usually called metallicity; this distinguishes elements produced by the Big Bang from those produced in stars), it is not as simple with AGNs. This is because, as many studies have shown (*e.g.* Bennert *et al.*, 2006; Nagao *et al.*, 2006), the results are not unique and their interpretation depends consequently on what is expected for these galaxies; for example, the common belief that all AGNs must have a high heavy element content, higher than solar, while in fact this is a result that applies only for Broad Line AGNs (Hamann and Ferland 1992).

But what if the problem is somewhere else? It is usually assumed that the variations observed in the spectra, which are the basis of the classification criteria in diagnostic diagrams, reflect only a difference in photo-ionization structures and parameters. Could these variations be also related to other physical differences, for example, related with the evolution and formation of the host galaxies (*e.g.* Pilyugin and Thuan 2011)? To explore further this hypothesis we have selected from SDSS the largest sample of NELGs with highest quality spectral data (161 577 galaxies) and extended the analysis based on standard diagnostic line ratios to include the chemical abundances, the morphologies of the hosts, their bulge masses and mean stellar population ages.

## METHOD

The data for our study come from the main catalogue of the Sloan Digital Sky Survey, Data Release 7 (SDSS DR 7; Abazajian *et al.*, 2009). We downloaded the spectra of 926 246 objects, already classified as galaxies and with narrow emission line features. We used the Virtual Observatory service at http://www.starlight.ufsc.br, where the spectra are already corrected for Galactic extinction, shifted to their rest frame, and re-sampled to $\Delta\lambda$ = 1 Å between 3400 and 8900 Å, ready to be processed by the spectral synthesis code STARLIGHT (Cid Fernandes *et al.*, 2005). STARLIGHT decomposes an observed spectrum in terms of a sum of simple stellar populations (SSPs), each of which contributes a fraction $x_j$ to the continuum flux at a chosen normalization wavelength ($\lambda_0$ = 4020 Å). We used a base of $N_{star}$ = 150 SSPs extracted from the models of Bruzual and Charlot (2003). The base components comprise 25 ages between $t_{star,j}$ = 1 Mega-years and 18 Giga-years, and six metallicities from $Z_{star,j}$ = 0.005 to 2.5 $Z_{solar}$ (Cid Fernandes *et al.*, 2007; Asari *et al.*, 2007). STARLIGHT produces automatically for each galaxy a stellar population template-corrected spectrum, which allows measuring with high precision some important attributes of the spectral lines, like the emission line fluxes, the equivalent width (EW) and full width at half maximum (FWHM). Through the fitted templates we can also retrieve the mean stellar population ages and stellar velocity dispersions of the galaxies.

By keeping only those galaxies with redshift z ≤ 0.25 that have line fluxes for the forbidden emission lines [OIII]λ5007, [NII]λ6584, [SII] λλ6717,6731 and the two Balmer lines Hα and Hβ with signal-to-noise ratios, S/N, greater or equal to 3 (S/N > 10 in the adjacent continuum), we reduce the sample to 161577 NELGs (only 17 % of the original sample). In figure 1, the standard diagnostic diagram based on the two line ratios I([OIII]λ5007)/I(Hβ) *vs.* I([NII]λ6584)/I(Hα) is used to determine the activity type of the galaxies in our sample. We count 89 941 SFGs (56 %), 46 880 TOs (29 %) and 24 756 AGNs (15 %). This diagram is insensitive to dust extinction (because the lines used in the ratios are affected by extinction by the same amount), and insensitive to the underlying continuum produced by the various stellar populations (because we have subtracted a stellar template adapted to each spectrum). Consequently, the uncertainties on the line ratios are small and do not affect our classification.

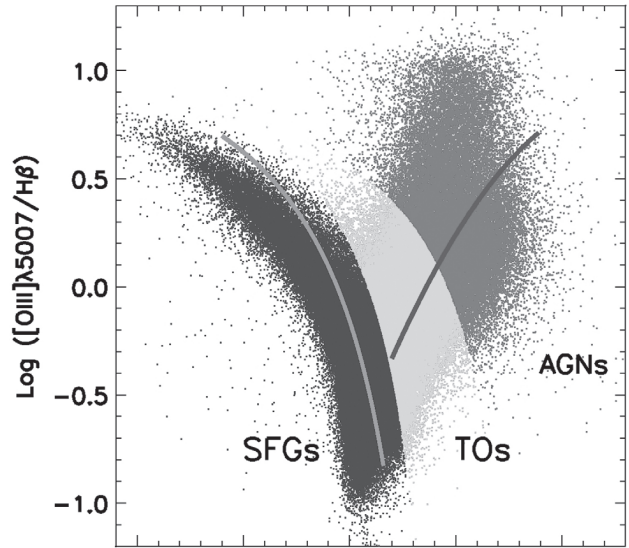

**Figure 1:** Standard diagnostic diagram for NELGS activity classification. The schemes applied to separate the NELGs into SFGs, TOs and AGNs, are those proposed by Kewley *et al.*, (2001) and Kauffmann *et al.*, (2003). Plot over the distribution of galaxies, we have traced the two sequences between [O/H] and [N/H]: the metallicity increase from up to down and the nitrogen abundance increase from left to right.

All the fluxes were corrected for dust extinction using the ratio AV produced by STARLIGHT. For the SFGs, the ratios I([OIII]λ5007)/I(Hβ) are transformed into O/H, using the empirical relation (Vacca and Conti 1992):

$$\log\left(\frac{O}{H}\right) = -0.69 \cdot (\log R_3) - 3.24 \quad (1)$$

where:

$$R_3 = 1.35 \times \frac{I([OIII]\lambda 5007)}{I(H\beta)} \quad (2)$$

The [O/H], is expressed using the standard scale units:

$$\left[\frac{O}{H}\right] = \left[12 + \log\left(\frac{O}{H}\right)\right] - 8.66 \quad (3)$$

where 8.66 is the metallicity of the Sun adopted on this scale (Asplund *et al.*, 2004). The abundance of nitrogen with respect to the oxygen, N/O, is determined following Thurston *et al.*, (1996). First we determine the temperature of the gas using the empirical relation:

$$T_{II} = 6065 + 1600 \cdot (\log R_{23}) + 1868 \cdot (\log R_{23})^2 + 2803 \cdot (\log R_{23})^3 \quad (4)$$

where:

$$R_{23} = \frac{I([OIII]\lambda 5007) + I([OIII]\lambda 4959) + I([OII]\lambda 3729) + I([OII]\lambda 3727)}{I(H\beta)} \quad (5)$$

Then we apply the empirical relation:

$$\log\left(\frac{N}{O}\right) = \log\left(\frac{I([NII]\lambda 6548) + I([NII]\lambda 6584)}{I([OII]\lambda 3727) + I([OII]\lambda 3729)}\right) + 0.307 - 0.02 \cdot \log(T_{II}) - \frac{0.726}{T_{II}} \quad (6)$$

where $T_{II}$ is in units of $10^4$ K.

The mean stellar age is determined following Asari *et al.*, (2007):

$$\log(t_{\text{star, j}})_L = \sum_{j=1}^{N_{\text{star}}} x_j \log(t_{\text{star, j}}) \qquad (7)$$

where the subscript *L* emphasizes that the mean is light-weighted.

Because of the finite size of the SDSS spectral fiber the stellar velocity dispersion, $\sigma_{\text{star}}$, deduced from the STARLIGHT template corresponds really to the dispersion of the stars in the inner part of the galaxies, which is dominated by the bulge. The bulge mass is thus determined by applying the virial relation to $\sigma_{\text{star}}$. We have verified that the range in redshifts of our sample do not produce spurious differences between the SFGs, TOs and AGNs.

The morphologies of the galaxies are determined according to a method which is based on the correlations that exist between the photometric colors, the inverse concentration index and the morphological types found by Shimasaku *et al.*, (2001) and Fukugita *et al.*, (2007). The *ugriz* colors are those of the SDSS photometric system: *u-g, g-r, r-i* and *i-z* (http://casjobs.sdss.org). The concentration index comes from SDSS and corresponds to $R_{50}/R_{90}$, the ratio of the Petrosian radii (Petrosian 1976) containing 50 % and 90 % of the total flux in the *r* band. A *K*-correction is applied to the Petrosian magnitudes using the code developed by Blanton and Roweis (2007). The morphologies were scaled using the index, T, which ranges from 0 to 6 (where 0 = E and 6 = Irr). The integer values (0 to 6) correspond to "integer morphological classes" and the half-integer values correspond to intermediate classes. The correspondence between the morphological indices and Hubble morphological types is presented in table 1.

**Table 1.**
Correspondence between Hubble morphology and morphological index, T.

| E | E/S0 | S0 | S0/Sa | Sa | Sab | Sb | Sbc | Sc | Scd | Sd | Sdm/Sm | Im |
|---|---|---|---|---|---|---|---|---|---|---|---|---|
| 0.0 | 0.5 | 1.0 | 1.5 | 2.0 | 2.5 | 3.0 | 3.5 | 4.0 | 4.5 | 5.0 | 5.5 | 6.0 |

The absolute B magnitudes were obtained using the Johnson-B band magnitudes synthesized from the SDSS magnitudes and the uncertainty on these values is of the order of 0.05 mag (Fukugita *et al.*, 1996) and used as a prior for the masses of the galaxies.

**DISCUSSION**

The diagnostic diagram presented in figure 1 contains 161 577 galaxies. With the exception of a few scattered points, the distribution of the data is seen to trace one continuous feature, under the form of the Greek letter v. This continuity in the distribution of the data suggests that it should be possible to explain the spectral characteristics of the NELGs assuming the variation of one predominant property. For the SFGs this property was already identified as the heavy element content (McCall *et al.*, 1985; Evans and Dopita 1985). In particular, the SFGs describe a sequence where the emission ratio [OIII]λ5007/Hβ is decreasing as the ratio O/H increases. The physical reason is well understood: the oxygen emission being the principal cooling mechanism in HII regions the higher the O/H ratio, the lower the temperature and the lower the ratio [OIII]λ5007/Hβ.

For the SFGs, it is also observed that as O/H increases, the ratio N/H increases (Thurston *et al.*, 1996; van Zee *et al.*, 1998). Therefore, we must recognize that the diagnostic diagram for the SFGs is also sensitive to the O/H and N/H abundance ratios of the galaxies.

Could this be also true for the TOs and AGNs? After all, the chemical elements in galaxies are a product of star formation, not of the SMBH. Could the chemical abundances also play an important role in determining the position of the AGNs in the diagnostic diagram of figure 1? The majority of the researchers accept that in the AGNs the emission-line ratio [NII]λ6584/Hα reflect an increase in N/H abundance ratio. Implicitly, they presume that the O/H ratio must also increase in the same direction. The possible physical justification is that nitrogen being dominantly a secondary element (in particular in the central region of massive galaxies) implies that N/O must increase linearly with O/H. But, then, how can we explain the fact that the ratio [OIII]λ5007/Hβ is increasing instead of decreasing like we observe in the SFGs? To explain this apparent contradiction it is further assumed that in the AGNs other ionization parameters—for example, increasing the ionizing parameter, increasing the electronic density, or assuming different values for the covering factor (Baskin *et al.*, 2005)—produce an increase of the temperature increasing significantly the [OIII]λ5007 ratios as O/H decreases. The problem with these solutions is that they depend on the models, and usually do not take into account other considerations like the chemical evolution of the galaxies or their formation process. The consistencies of these solutions are usually not independently verified.

Assuming that the ratio O/H in AGNs continues to increase with the nitrogen emission produces still another problem. It suggests the presence of two scales for O/H on the [OIII]λ5007/Hβ axis. One for the SFGs and another for the AGNs, and at some point of the diagnostic diagram the O/H scale must change allowing [OIII]λ5007/Hβ to increase. Consequently, some authors have proposed to use [NII]λ6584/Hα as the indicator of the heavy element content for the AGNs. But how to calibrate this relation is not clear, and except for the unverified assumption of a secondary relation for nitrogen applying to all galaxies, the external consistency of this solution, again, has not been fully tested.

Besides, we also have the problem that the spectral criteria adopted to classify the NELGs do not allow separating them in an absolute way. The empirical limits between the SFGs and TOs, and between the TOs and AGNs, are always smeared by numerous galaxies with intermediate spectral characteristics. The hypothesis of two scales also seems in contradiction with the TOs which are the archetype of objects having a double nature, as a mixture of SFGs and AGNs.

There is one alternative, which is to consider only one scale for the relation between O/H and [OIII]λ5007/Hβ. This alternative suggests the cooling of the gas by oxygen emission is still the dominant effect in the narrow line regions of AGNs. In Figure 1, we have fitted two empirical relations between [OIII]λ5007/Hβ and [NII]λ6584/Hα. Comparing with CLOUDY models obtained by Bennert *et al.*, (2006) on galaxies similar to the TOs we have found that, as a first approximation, a consistent calibration in the ratio O/H with an uncertainty of 0.2 dex (based on the dispersion) could be obtained using the same relation with [OIII]λ5007/Hβ as for the SFGs (Coziol *et al.*, 2011). This model produces two consistent relations for the increase of N/H as a function of O/H: on the SFGs side the N/H grows with O/H, while on the TOs and AGNs the N/H increases as the ratio O/H decreases. That is, our model predicts a growing excess in the ratio N/H relative to O/H passing from the TOs to the AGNs (Osterbrock 1970; Storchi-Bergmann and Pastoriza 1989; Storchi-Bergmann 1991; Hamann *et al.*, 1993).

**The origin of the excess of Nitrogen in galaxies**

Concentrating on the SFGs for which the estimation of the ratio O/H is more secured, we show in figure 2 the results of two ionization models produced using CLOUDY (see Coziol *et al.*, 1999 for details), where the abundance of nitrogen is the only factor that is varying—increasing by 0.2 dex in the Seq. 2 model compared to that in the Seq. 1 model. We can see that most of the SFGs fall between these two limits. Using the median, we have separated the SFGs in two groups. We identify the galaxies with higher N/H abundance ratio as the Starburst Nucleus Galaxies (SBNGs).

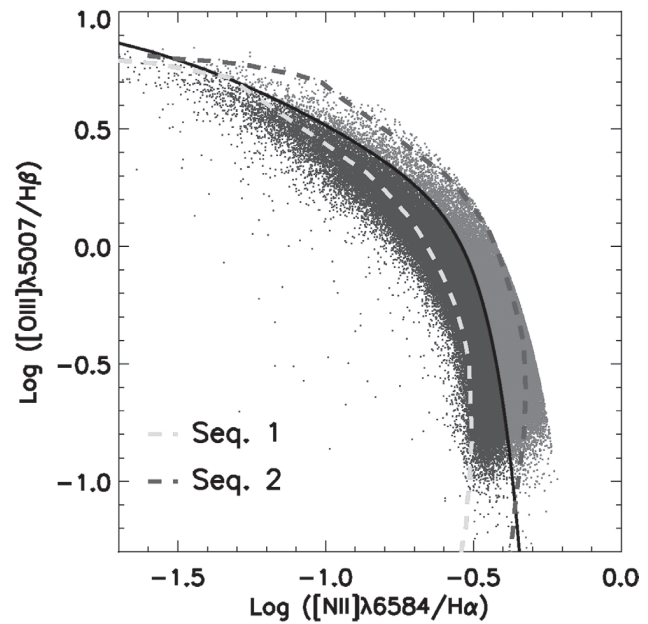

**Figure 2:** Diagnostic diagram for the SFGs only. The two sequences (Seq. 1 and Seq. 2) are results of CLOUDY modelisation, where the abundance of nitrogen in increased by 0.2dex while all the other parameters are kept constant; Seq. 2 model has an excess of nitrogen compared to Seq. 1 (Coziol et al. 1999).

In figure 3, we show a second diagnostic diagram based on the ratios [SII]λλ6717,6731/Hα *vs.* [NII]λ6584/Hα. The TOs, showing an excess of excitation in [SII]λλ6717,6731/Hα and [NII]λ6584/Hα, are found on the higher-right side of this diagram (Coziol *et al.*, 1999). Note how the distinction between the SFGs and SBNGs seems to more clearly isolate the SFGs in the lower-left part of the diagram. However, the separation is not complete, as already mentioned. The same is also true comparing the SBNGs with the TOs.

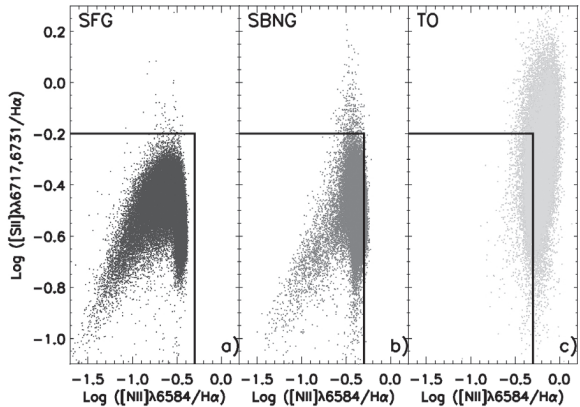

**Figure 3:** Spectral diagnostic diagram separating AGNs from SFGs (Coziol et al. 1999). The two limits allow identifying galaxies with an excess of emission, which are found mostly in the upper-right part of the diagram.

To discuss the difference in terms of the abundances, we have traced in figure 4 and in figure 5 the same diagrams but as a function of the ratio [O/H] relative to the Sun, using our empirical calibration for the TOs. In figure 4 it is observed that the SFGs are showing an inversion in the relation between [O/H] and the emission ratio [SII]$\lambda\lambda$6717,6731/H$\alpha$: at first the ratio increases with [O/H], but then it decreases (mostly after [O/H] > 0). The SBNGs on the other hand are almost only showing the inverse correlation, the ratio [SII]$\lambda\lambda$6717,6731/H$\alpha$ decreasing with [O/H]. Remarkably the TOs show almost no relation between [O/H] and the emission in sulfur. The results of the Spearman and Pearson correlation tests can be found in table 2, where the tests was done using only the galaxies with [O/H] > 0. We measure a weak negative correlation in the SFGS and TOs, and a stronger one in the SBNGs.

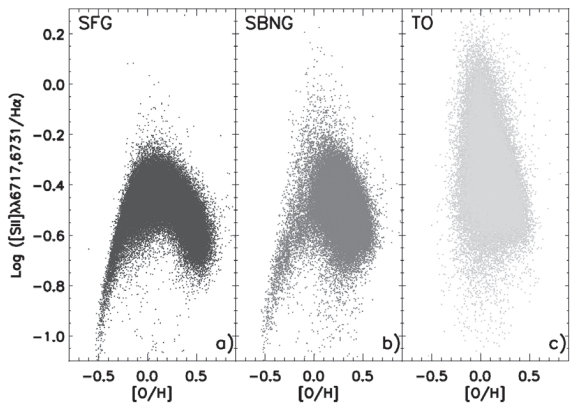

**Figure 4:** Variation of emission in sulfur with the metallicity. The SFGs show an inversion of the relation between the sulfur emission and [O/H] and the TOs show almost no relation.

**Table 2.**
Correspondence between Hubble morphology and morphological index, T.

|  | Spearman |  | Pearson |  |
|---|---|---|---|---|
| [NII]$\lambda$6584/H$\alpha$ vs [O/H] | $r_s$ | $P(r_s)$ | $r$ | $P(r)$ |
| SFG | 0.522 | 2.2e-16 | 0.579 | 2.2e-16 |
| SBNG | 0.820 | 2.2e-16 | 0.790 | 2.2e-16 |
| TO | -0.059 | 2.2e-16 | -0.077 | 2.2e-16 |
| [SII]$\lambda\lambda$6717,6731/H$\alpha$ vs [O/H] | $r_s$ | $P(r_s)$ | $r$ | $P(r)$ |
| SFG | -0.401 | 2.2e-16 | -0.377 | 2.2e-16 |
| SBNG | -0.619 | 2.2e-16 | -0.596 | 2.2e-16 |
| TO | -0.354 | 2.2e-16 | -0.354 | 2.2e-16 |

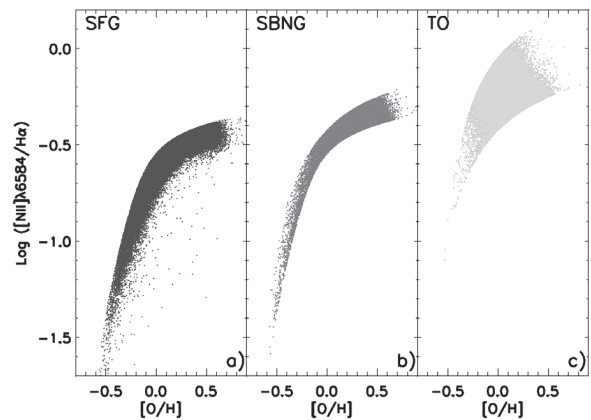

**Figure 5:** Variation of nitrogen emission with [O/H].

In figure 5 we find no inversion with [O/H] for the nitrogen emission, but rather saturation at high [O/H] values. For the SFGs, exactly at the same value in [O/H] where the increase in nitrogen emission begin to saturate, we observe the inversion for the sulfur emission. In the SBNGs, the nitrogen emission rises higher than in the SFGs and seems to continue to rise, although more slowly, at high [O/H]. Of extreme importance, we note for the TOs that our [O/H] calibration still predicts an increase of nitrogen emission with [O/H]. However, the spread in nitrogen emission is larger than for the SFGs and SBNGs, which suggests that other physical parameters also influence the intensity of the emission lines in these objects. This is confirmed by the Spearman and Pearson correlation tests in table 2. We find a strong correlation for the SBNGs, a weaker one for the SFGs and no correlation for the TOs.

In our previous discussion about the O/H calibration in the diagnostic diagram based on [OIII]$\lambda$5007/H$\beta$ we have claimed that no inversion was expected for the O/H relation for the AGNs. But, here we are observing some sort of inversion for the sulfur lines, with the difference that this inversion seems to appear in the sample of SFGs not in the AGNs. Note also that the inversion would not be in the sense as suggested before, because the sulfur emission would suggest O/H must decrease not increase. An explanation in terms of the abundance of sulfur is more plausible. Considering that sulfur is produced by massive OB stars in young HII regions, the inversion suggests the SBNGs have already stopped forming this element in great quantity. This implies that the SBNGs follow a different chemical evolution than the SFGs. This is illustrated in figure 6, where we compare the abundance in nitrogen (N/O) as a function of [O/H] (here we have applied the same method to determine N/O to the TOs, assuming their SFGs components produced the abundances). Overplotted on the figures, we trace the results of different chemical evolutionary models (*secondary* and *primary + secondary* production of nitrogen) as proposed by Vila-Costas and Edmunds (1993). In general, the SFGs seem to follow relatively well the model for a *secondary* production of nitrogen. The SBNGs on the other hand do not follow this model. For the same values in O/H they show an excess of nitrogen compared to the SFGs. In fact, we observe a non continuous behavior in the increase of nitrogen with O/H: there is a sharp jump by 0.3 dex in nitrogen at log (O/H) = - 3.4. Exactly the same phenomenon was observed before by Coziol *et al.*, (1999) in a comparatively very small sample of SBNGs (a few hundreds data points). The fact that we observe exactly the same phenomenon for a sample of 33 929 galaxies suggests that this behavior seems like a common trait of these galaxies.

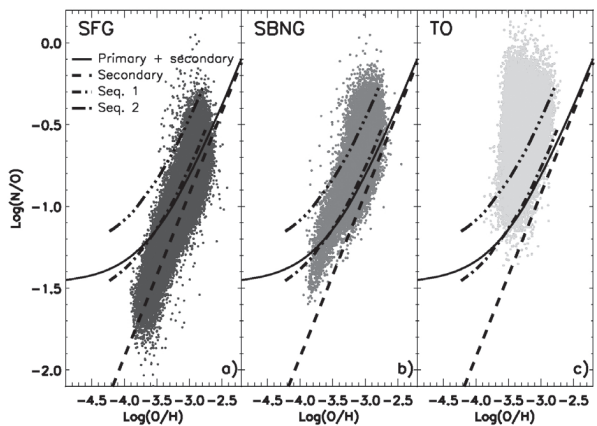

**Figure 6:** The abundance of nitrogen as a function of the O/H ratio. The secondary (dash line) and primary + secondary (continuous line) relations are also shown together with Seq 1 (dot-dash line) and 2 (3 dot-dash line). The SBNGs do not follow the secondary relation. There is a distinct jump of log(N/O) = -1.3 at about log(O/H) = − 3.4. Exactly the same behavior was observed before by Coziol *et al.*, (1999), with a hundred points in their diagram, while here we have more than 33 929 individual galaxies.

In Coziol *et al.*, (1999) the model of Garnet (1990) was used to explain that an excess in N/O could result from sequences of bursts over a short period of about a Giga-year. This sequence of bursts was assumed to be connected with the bulge formation process of these galaxies. Our approximation for the TOs suggests they may have gained a higher excess in nitrogen than the SBNGS because they have built (in the same way as the SBNGs) more massive bulges. Based on the SBNGs, we may have found therefore a justification for an excess of N/O in the AGNs: assuming they formed in the same way as the SBNGs, they would all have an excess of nitrogen due to their massive bulges.

**Relations with physical parameters of the galaxy hosts**

If our interpretation for the differences in nitrogen abundance of the SFGs is correct we would thus expect these differences to be connected with different physical properties of the host galaxies. In figure 7 we compare the mean stellar ages, the morphologies, the bulge masses and absolute blue magnitude of the galaxies. All the differences observed were tested for statistical significance using a new parametric test, the t-max test, already implemented in R (see Torres-Papaqui *et al.*, 2011 for explanations and details). The results are reported in Table 3. From the upper left to the lower right panels we see that the TOs have the oldest stellar populations, by almost one order of magnitude. We find also that the morphologies are significantly different, with the TOs being encountered in earlier-type galaxies than the SBNGs. Consequently, the TOs also show the most massive bulges, in good agreement with what we expect based on our hypothesis for the differences in abundance. On the other hand, these differences in bulge mass and morphology do not seem to have anything to do with the total masses of the galaxies, as can be judged from the absolute blue magnitudes. In particular, no obvious difference is observed between the SBNGs and TOs (although the statistical tests suggest the TOs are less massive). In general, the differences between all these galaxies are not related to their total masses, but on how the masses are distributed, a higher number of stars being under the form of massive bulges in the TOs than in the SBNGs. Our model predicts that for the same O/H ratio the TOs have a higher excess of nitrogen than the SBNGs as shown in the lower center and right panels. All these differences can be explained in the hierarchical structure formation theory using the multiple mergers model.

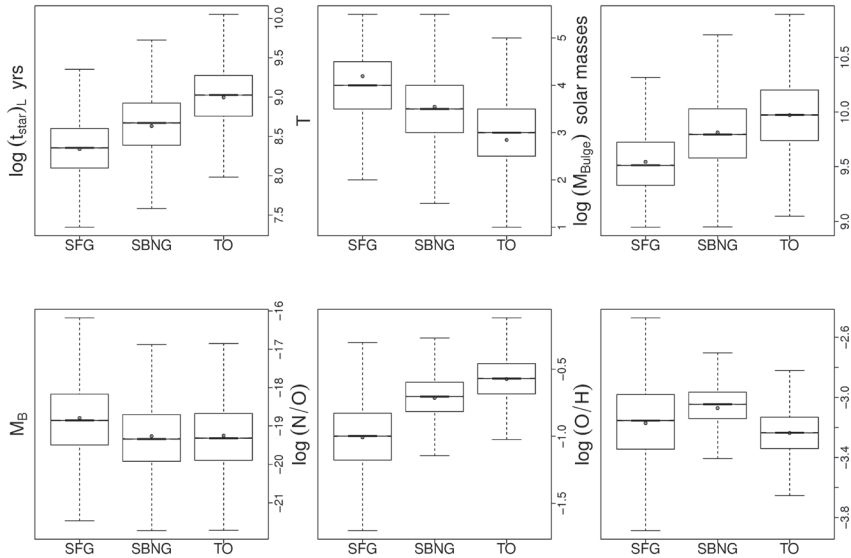

**Figure 7:** Box-whisker plots comparing the basic characteristics of the SFGs, SBNGs and TOs. The significance of the statistical tests is shown in Table 3.

**Table 3.**
Multiple comparisons of means with Tukey contrasts

| Pairs | Significance for $Pr(>|t|)$ | | | | | |
|---|---|---|---|---|---|---|
| | $Log (t_{star})L$ | T | $Log (M_{Bulge})$ | MB | $Log (N/O)$ | $Log (O/H)$ |
| SBNG - SFG | *** | *** | *** | *** | *** | *** |
| TO - SFG | *** | *** | *** | *** | *** | *** |
| TO - SBNG | *** | *** | *** | *** | *** | *** |

95 % confidence level: *** for Pr < 0.001, ** for Pr < 0.01, * for Pr < 0.05, . for Pr < 0.1 and nothing for 1.

### The multiple mergers model for SBNGs and TOs

Our calibration for the abundance of the SFGs and AGNs is consistent with different formation processes for the galaxies. Our interpretation of these observations can be found in figure 8. There are many structural similarities between the bulges of spiral galaxies and elliptical galaxies, suggesting similar formation mechanisms (*e.g.* Jablonka *et al.*, 1996). In particular, elliptical galaxies are known to have endured higher astration rates than spiral galaxies when they formed (Sandage 1986): they transformed almost all their gas into stars in a very short period of time. Assuming galaxies form by a succession of star forming episodes—an assumption necessary to produce the galaxy mass-[O/H] relation (Coziol *et al.*, 1998)—galaxies with high astration rates would thus be expected to have formed most of their stars in the past, given that the reservoir of gas is limited and the galaxy would have consumed it rapidly and stopped forming stars relatively early. This would also explain the differences in mean ages between the SFGs, SBNGs and TOs. This difference in stellar population age is also consistent with an excess of nitrogen in the TOs and SBNGs compared to the SFGs, because a higher number of low mass stars produced in these galaxies would have had time to evolve.

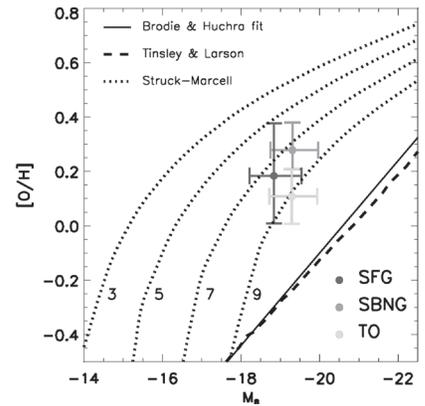

**Figure 8:** The chemical evolution of galaxies related with their formation process. The continuous curve is the Brodie-Huchra relation for elliptical galaxies. Also shown are two models of multiple mergers: one by Tinsley and Larson (1979), which explains the Brodie-Huchra [O/H]-M¬B relation for the elliptical galaxies, and the modified version proposed by Struck-Marcell (1981) to explain the chemical evolution of the spiral galaxies. The numbers indicate the approximate number of mergers necessary in the models to reproduce the [O/H] observed values.

The difference in heavy element content between the TOs and SBNGs is explained by the variations introduced by Struck-Marcell (1981) in the chemical evolution scenarios based on multiple mergers proposed by Tinsley and Larson (1970). In the hierarchical structure formation model, the galaxies form by a succession of mergers each time accompanied by a burst of star formation. Struck-Marcell (1981) shown that by allowing a higher amount of gas than stars to be accreted during each merger, the heavy element content is seen to increase more rapidly. This is because each time stars are formed, a higher fraction of enriched gas goes into their formation.

There would be therefore two merger scenarios for the formation of galaxies: the dry merger scenario as proposed by Tinsley and Larson (1980), and which explains the formation of bulge-dominated galaxies (like the TOs and AGNs), and the wet scenario as proposed by Struck-Marcell (1981) which explains galaxies dominated by a disk (SFGs and SBNGs).

The key factor in determining the morphologies of galaxies is the collapse time (Sandage 1986). The dry merger scenario corresponds consequently to the case where we have a high number of mergers over a short period of time, while in the wet merger model a lower frequency of mergers is implied. Because each merger produces a burst in star formation, galaxies following the dry merger model have high astration rates which produce massive bulges.

**CONCLUSIONS**

We have shown that the diagnostic diagrams used to classify the NELGs according to their nuclear activity are also sensitive to the chemical abundances. Furthermore, we have shown that the chemical abundances are correlated with the physical characteristics of the galaxies, like their mean stellar population ages, their morphologies and the masses of their bulges. All these correlations are consistent with different formation processes for the galaxies. More specifically, it suggests that galaxies that form a massive bulge produce an excess of nitrogen. If the AGNs form the same way as the SBNGs, then they would also have an excess in nitrogen, possibly proportional to the mass of their bulges. The standard spectral diagnostic diagrams for the NELGs looks like snapshot images of a population of galaxies that formed in different initial conditions.

**ACKNOWLEDGMENTS**

J.P. Torres-Papaqui acknowledges PROMEP for support grant 103.5-10-4684.